\documentclass[published]{JHEP3} 
\JHEP{00(2005)000}
\JHEPspecialurl{http://jhep.sissa.it/JOURNAL/JHEP3.tar.gz}
\usepackage{epsfig,multicol,subfigure}
\newcommand\fverb{\setbox\pippobox=\hbox\bgroup\verb}
\newcommand\fverbdo{\egroup\medskip\noindent%
                        \fbox{\unhbox\pippobox}\ }
\newcommand\fverbit{\egroup\item[\fbox{\unhbox\pippobox}]}
\newbox\pippobox
\def\be{\begin{eqnarray}}
\def\ee{\end{eqnarray}}
\def\beq{\begin{equation}}
\def\eeq{\end{equation}}
\def\ba{\begin{array}}
\def\ea{\end{array}}
\def\hc{{\rm H.c.}}
\def\bec{\begin{center}}
\def\ec{\end{center}}

\title{Partially composite two-Higgs doublet model}

\author{Byungchul Chung,~Kang Young Lee \\
        Department of Physics, KAIST, Daejeon 305-701, Korea \\
        E-mail: \email{crash@muon.kaist.ac.kr},
                ~\email{kylee@muon.kaist.ac.kr}}

\author{Dong-Won Jung,~Pyungwon Ko \\
        School of Physics, KIAS, Seoul 130-722, Korea \\
        E-mail: \email{dwjung@kias.re.kr}, 
               ~\email{pko@kias.re.kr} }

\received{} 
\revised{}
\accepted{}           

\preprint{KIAS--P05038}     

\abstract{
In the extra dimensional scenarios with gauge fields in the bulk, 
the Kaluza-Klein (KK) gauge  bosons can induce Nambu-Jona-Lasinio (NJL) 
type attractive four-fermion interactions, which can break electroweak 
symmetry dynamically with accompanying composite Higgs fields.  
We consider a possibility that electroweak symmetry breaking (EWSB) 
is triggered by both a fundamental Higgs and a composite Higgs 
arising in a dynamical symmetry breaking mechanism 
induced by a new strong dynamics.
The resulting Higgs sector is a partially composite two-Higgs doublet 
model with 
specific boundary conditions on the coupling and mass parameters originating
at a compositeness scale $\Lambda$. 
The phenomenology of this model is discussed including 
the collider phenomenology at LHC and ILC.
 }

\keywords{dynamical symmetry breaking, two-Higgs doublets model}

\dedicated{}

\begin{document}


\section{Introduction}

Understanding the origin of electroweak symmetry breaking (EWSB) is 
one of the most important problems in particle physics. 
EWSB is strongly tied with the masses of chiral fermions and electroweak 
gauge bosons as well as CP violation in the standard model (SM) of 
Glashow-Salam-Weinberg.
The nature of EWSB will be experimentally studied in detail at the CERN 
Large Hadron Collider (LHC) and the future $e^+ e^-$ linear collider (ILC).  
One may observe the force quanta, Higgs boson(s), that breaks electroweak 
symmetry in the standard model (SM), or its supersymmetric extensions. 
Or one may observe another signatures if there is no fundamental Higgs. 
It is an important phenomenological issue to extract maximal informations 
on the EWSB sector from the data obtainable at the LHC and the ILC. 
On the theoretical side, one has to study various models of EWSB including 
the SM which has a single Higgs doublet, compare them with the data, and 
figure out which mechanism of EWSB is realized in nature. 

In fact, there have been a lot of efforts to construct interesting 
models for the EWSB for more than the last two decades.
In principle, studying EWSB is not necessarily related to solving 
the gauge hierarchy problem which is a fine tuning problem. 
But they are often interwined in actual model buildings. 
We would not care about solving the gauge hierarchy problem.  Then 
there are a few different avenues to extend the SM regarding the EWSB:

\begin{itemize}

\item More fundamental Higgs : 
More fundamental Higgs singlets, doublets and triplets. 
This includes the general 2 Higgs doublet model, the minimal 
supersymmetric standard model (MSSM) and its various extensions
\cite{higgshunter}.  In most cases, theories are weakly coupled and 
perturbation theory works.  

\item Composite Higgs or dynamical EWSB : 
(extended) Technicolor scenarios,
top condensate, topcolor, top seesaw, Little Higgs etc.
\cite{hillsimmons,BHL,miransky,topcolor,topseesaw}.
In most cases, the EWSB sector is strongly coupled, and perturbation 
theory cannot be applied directly. One often has to construct
effective field theory (EFT) to analyze the EWSB sector. 
There could be composite Higgs in the low energy spectra, but not 
necessariliy or always.

\item Extra dimension : 
More options are avaiable in the extra dimensional scenarios. 
EWSB can arise either by fundamental Higgs, or by boundary conditions 
or by dynamical symmetry breaking from extra dimensional QCD 
\cite{koba,dobrescu1,dobrescu2}. 
In fact, all the three options are generic in the extra dimensional 
scenarios and should be considered altogether in principle. 
This is in a sense similar to supersymmetry (SUSY) breaking scenarios, 
gravity mediation, gauge mediaion, anomaly mediation, gaugino mediation, 
etc., and some of these mediations could be comparable with each other. 

\end{itemize}

Dynamical symmetry breaking \'{a} la Bardeen-Hill-Lindner (BHL)
\cite{BHL} is a particularly interesting scenario, 
since the heavy top mass is intimately related with a new strong dynamics 
that condenses the $t\bar{t}$ bilinear, 
and breaks the EW symmetry down to $U(1)_{\rm EM}$.
Both heavy top mass and Higgs mass are generated dynamically, in anology 
with superconductivity of Bardeen-Cooper-Schrieffer (BCS) \cite{bcs}. 
The original version of BHL with 3 families predicts that 
the top mass should be larger
than $\sim 200$ GeV, which is no longer viable considering 
the present measurement $m_t = 178 \pm 4.3$ GeV \cite{topmass}.
Extension of BHL with two Higgs doublets has a similar shortcoming 
\cite{luty}. The top mass can be lowered to the observed value, 
if one considers supersymmetric extension \cite{carena} or if there is
the 4-th generation \cite{luty}.
Another conceptual problem of BHL scenario is the origin of the new strong
interactions that triggers electroweak symmetry breaking. 
The attractive 4-fermion interaction is simply put in by hand 
within the BHL model. 
Despite these drawbacks, we believe that it is worthwhile to consider
variations of the BHL scenario, since these two drawbacks can be easily
evaded within extra dimensional scenarios, without ruining its niceties: 
this scenario is attractive, could be generic, 
and provides dynamical origins for fermion and gauge boson masses 
(at least a part of them) in the extra dimensional scenarios. 

A new trend in model building for the last decade was to use extra 
dimesions to solve gauge hierarchy problem, and/or fermion mass hierarchy
\cite{extraD}.
If QCD is a bulk theory, then it is possible that the extra dimensional QCD
can induce attractive Nambu-Jona-Lasinio (NJL) type four-fermion interaction 
in the low energy regime, and dynamical symmetry breaking can occur as usual. 
In short, one has a natural setting for dynamical symmetry breaking 
from extra dimensional QCD. 
It should be emphasized that this is completely different from another 
popular way of symmetry breaking in extra dimension, namely symmetry 
breaking by boundary conditions. 
We believe that it is not an option but an obligation to study 
the dynamical EWSB in extra dimensional scenarios, 
if once we have gauge theories in the bulk. 
After the large extra dimension scenario was put forward,
a few groups studied dynamical symmetry breaking in flat extra dimensions
\cite{DSBlarge} and in warped extra dimension \cite{DSBwarped}. 
The qualitative result from these studies is that 
it is indeed possible to have dynamical EWSB from extra dimensional QCD. 
Then in extra dimensional scenarios, electroweak symmetry 
can be broken by fundamental Higgs, by boundary condition or 
by some dynamical mechanism. 
Generically all three possibilities could be present altogether. 
In recent literatures, each option of EWSB in extra dimensional scenarios 
was discussed in detail. In most cases, the gauge symmetries are 
broken by both fundamental Higgs as well as   
the nontrivial boundary conditions, or completely by boundary conditions 
without Higgs. Another interesting possibility that electroweak symmetry 
is broken by fundamental Higgs VEV's, and dynamically by $t\bar{t}$ 
condensate as well.  It is the purpose of this work to consider this 
possibility in a minimal setup \footnote{Some works have been done on related
topics by several different groups, {\it i.e. },
a technicolor with heavy scalar doublet
 \cite{hscalar} and  bosonic topcolor (technicolor) \cite{bosontop}.}.

 In this paper, we consider an extension of BHL scenario, where 
one has a fundamental Higgs from the beginning, and the $t\bar{t}$ 
bilinear condensate due to a new strong interaction which is triggered by
extra dimensional QCD.
We assume this is achieved by embedding the SM in the higher dimensional
spacetime with an appropriate extra dimensional QCD.
Since we have both fundamental and composite Higgs fields,
it is natural that the low-energy effective theory is the two-Higgs
doublet model.  If we assume that the fundamental Higgs couples 
only to the bottom quark and the top quark purely receives its 
large mass from dynamical EWSB as in BHL,  then the resulting low 
energy effective theory is a Type-II two-Higgs doublet model 
with one fundamental and one composite  Higgs doublets. 
In our model, the top mass can be fit to the observed value 
for the limited range of $\tan\beta $. 
Actually only a narrow window for $0.45 \lesssim \tan\beta \lesssim 1$ 
is allowed in our model for a given compositeness scale $\Lambda$. 
For such a $\tan\beta$, roughly half of the $W$ and $Z$ masses come 
from the fundamental Higgs and half from the dynamical symmetry breaking. 
Since the phyical Higgs bosons are linear combinations of fundamental and 
composite Higgs bosons, we call it the partially composite Higgs boson. 
In a sense, our model is somewhere between the general two Higgs doublet 
model and the composite two Higgs doublet model by Luty \cite{luty}.
Compared with the model by Luty, we have one more parameter, 
the Yukawa coupling between the fundamental Higgs $\phi$ and 
the bottom quark at the compositeness scale.  
Then we can fit the heavy top mass without trouble unlike the Luty model. 
So the phenomenological disaster of the BHL-like top condensation model 
is gone in our model. 

Both in the BHL model and our model, 
the fine tuning problem of the Higgs mass is not explained.  
Unlike the general two Higgs doublet model, this model has only two more 
free parameters compared with the SM: the CP-odd neutral Higgs mass 
$m_A$ and the composite scale $\Lambda$.  (The quartic self-coupling 
$\lambda_{10}$ in our model is a free parameter as in the SM.)  
Therefore our model is more predictable and testable than the general 
two-Higgs doublet models, and could be possibly  verified or excluded 
in the future colliders. 

Although our work is motivated by the extra dimensional scenarios, 
the model presented in this paper is not a genuine model that would be
obtained when we embed the SM in the higher dimensional world. 
Generically we would have Type-III general two-Higgs doublet models if 
only the top quark couples to the dynamical symmetry breaking sector.
If the bottom quarks also couple to the dynamicall symmetry breaking 
sector, then the resulting low energy effective theory would be 
a three-Higgs doublet model. In this work, we assume that it is the top
quark which feels the dynamical EWSB  sector, and  we 
will introduce a discrete symmetry in order to reduce 
the Type-III Higgs doublet model to the Type-II  Higgs doublet model.
This will help to suppress dangerous Higgs mediated flavor changing 
neutral current processes. And the top mass is generated entirely from 
dynamical EWSB. 
It is beyond the scope of the current paper to discuss more general and 
realistic models with 3 generations of fermions into account, 
since it requires a more involved RG analysis for general two 
Higgs doublet model. 
This issue will be addressed in the future publication \cite{work1}.

This paper is organized as follows. In Section 2, we define our model
extending the BHL scenario. In Section 3, we present 
the mass spectra and the couplings of two-Higgs doublets in our model. 
In Section 4, we consider the Higgs boson productions at ILC and LHC,
and study the discoversy potential therein. 
We use the one-loop RG equations for couplings in the effective theory 
of our model, which are collected   in Appendix for convenience.


\section{A Model of Dynamical Symmetry Breaking with
a Fundamental Scalar}

We introduce a strong dynamics to the standard model at some  
high scale  $\Lambda$, which is  effectively described by
the NJL type four-fermion interaction term. 
Although we don't have to specify the origin of this NJL type interaction, 
we have in mind the KK gauge boson exchange as the origin of this new strong
interaction as discussed in the Introduction. 
As a minimal extension of the SM, we assume that this new strong dynamics 
acts only on top quark. 
Then we can  write the lagrangian at the scale $\Lambda$ as:
\be
{\cal L} = {\cal L}_{\rm SM} +
      G ( \overline{\psi}_{L} t_{R} )( \overline{t}_{R} \psi_{L} ),
\ee
where
\be
{\cal L}_{\rm SM} = {\cal L}_{\rm gauge} + {\cal L}_f + {\cal L}_{\phi}
            + (y_{t0}~\overline{\psi}_L t_R \tilde{\phi} + \hc)
            + (y_{b0}~\overline{\psi}_L b_R \phi + \hc) 
\ee
and $\tilde{\phi} \equiv i \sigma_y \phi^*$. 
The fermion and scalar field lagrangians are gauge invariant kinetic 
terms, given by
\be
{\cal L}_f &=& \bar{\psi}_L^a ~i \!\!\not \!\!D \psi_L^a 
                        + \bar q_R^a ~i \!\!\not \!\!D q_R^a,
\nonumber \\
{\cal L}_{\phi} &=& (D_\mu \phi)^\dagger(D^\mu \phi) - V(\phi),
\ee
where $a$ is the generation index, $\psi_L$ is the SU(2)$_{\rm L}$
doublet and $q_R$ the SU(2)$_{\rm L}$ singlet.
The scalar potential is given by
\be
V(\phi) = m_0^2 \phi^\dagger \phi
          + \frac{1}{2} \lambda_0 (\phi^\dagger \phi)^2 
\ee
as in the SM. $y_{t0}$ and $y_{b0}$ are Yukawa couplings of the top and
the bottom quarks to the fundamental Higgs field $\phi$.  
The Yukawa couplings for the 1st and the 2nd generations 
do not play any roles in our analysis, 
and will be ignored  in the following.

Introducing an auxiliary scalar doublet field $\Phi (x)$,
we can rewrite the NJL term in Eq.(2.1) as
\be
{\cal L} = {\cal L}_{\rm SM}
           + g_{t0} ( \overline{\psi}_{L} t_{R} \tilde \Phi + \hc )
           -M^2 \Phi^{\dagger} \Phi,
\ee
where $G = g_{t0}^2 / M^2$ and $g_{t0}$ is a newly defined Yukawa coupling
between the top quark and the auxiliary scalar field $\Phi$.
The mass scale $M$ will be generically of order $\Lambda$. 
The new scalar field $\Phi$ describes the composite scalar bosons that appear 
when the $\langle \bar{t} t \rangle$ develops nonvanishing VEV and breaks 
the electroweak symmetry by NJL type new strong dynamics. 
Then we have one fundamental scalar field $\phi$ and
one composite scalar field $\Phi$, although $\Phi$ is not a dynamical
field at the scale $\Lambda$. Far below the scale $\Lambda$, 
the $\Phi$ field will develop the kinetic term due to quantum corrections  
and become dynamical. The resulting low energy effective field theory will
be two-Higgs doublet model, one being a fundamental Higgs $\phi$ and the 
other being a composite Higgs $\Phi$.  
Thus it can be called a partially composite two-Higgs  doublet (PC2HD) model.  
In a general two-Higgs doublet model, 
one may have too excessive FCNC amplitudes mediated by neutral 
Higgs bosons. This phenomenological problem can be avoided if one invokes 
the Glashow-Weinberg criteria \cite{weinberg}.
For simplicity, we assign a $Z_2$ discrete symmetry under which 
the lagrangian is invariant ;
\begin{eqnarray*}
( \Phi,~\psi_L,~U_R ) & \to & + ( \Phi,~\psi_L,~U_R ) , \\
( \phi , D_R ) & \to & - ( \phi, D_R ) .
\end{eqnarray*}
With this $Z_2$ discrete symmetry, the Yukawa term 
$(y_{t0}~\overline{\psi}_L t_R \phi + \hc)$ is forbidden
in the lagrangian of Eq. (2.2), and only the $y_{b0}$ coupling term remains
\footnote{From now on, we will rename $y_{b0}$ as $g_{b0}$.}.  
In consequence, our model becomes the Type-II two-Higgs doublet model 
as the minimal supersymmetric standard model (MSSM),  
where the fundamental scalar field $\phi$ couples to the down-type quarks, 
and the composite scalar $\Phi$  couples to the up-type quarks.
The Higgs mediated FCNC is naturally suppressed by construction 
\footnote{If one allowed the fundamental Higgs to couple to the top quark, 
the top mass would get contributions both from the $\langle \phi \rangle$
and $\langle \Phi \rangle$, and the resulting low energy effective theory
will be a Type-III two-Higgs doublet model. 
We postpone discussing this case for the future study \cite{work1}. }.

We write the effective lagrangian far below $\Lambda$ as
\be
{\cal L} &=& {\cal L}_{\rm gauge} + {\cal L}_f
               + (D_\mu \phi)^\dagger(D^\mu \phi)
               + (D_\mu \Phi)^\dagger(D^\mu \Phi)
               \nonumber \\
               &&~~~~~~
               + (g_b \overline{\psi}_{L}b_R \phi + {\rm H.c})
               + (g_t \overline{\psi}_{L}t_R \tilde \Phi + {\rm H.c})
               - V(\phi, \Phi),
\ee
where the most general Higgs potential is given by
\be
V(\phi, \Phi) &=& \mu_1^2\phi^\dagger\phi + \mu_2^2\Phi^\dagger \Phi
              + ( \mu_{12}^2  \phi^\dagger \Phi + \hc)
\nonumber \\
         &&~+ \frac{1}{2}\lambda_1(\phi^\dagger\phi)^2
              + \frac{1}{2}\lambda_2(\Phi^\dagger \Phi)^2
\nonumber \\
         &&~+ \lambda_3(\phi^\dagger\phi)(\Phi^\dagger \Phi)
              + \lambda_4|\phi^\dagger \Phi|^2
              + \frac{1}{2} [ \lambda_5 (\phi^\dagger \Phi)^2 + \hc ].
\ee
In the scalar potential, we have introduced a dimension-two  
$\mu_{12}^2$ term that breaks the discrete symmetry softly  
in order to generate the nonzero mass for the CP-odd Higgs boson. 
Otherwise the CP-odd Higgs boson $A$ would be an unwanted axion related
with spontanesously broken global $U(1)$ Peccei-Quinn symmetry, 
which  would be a phenomenological disaster.
This $\mu_{12}^2$ term will not induce dangerously large FCNC
amplitudes at loop levels, thus does not spoil our original motivation to 
consider the discrete symmetry \cite{higgshunter}. 
In the next section, this $\mu_{12}^2$ parameter will be traded with the 
$m_A^2$, the (mass)$^2$ parameter of the CP-odd Higgs boson, 
which is another free parameter of our model.

The renormalized lagrangian for the scalar fields at low energy 
is given by
\be
{\cal L}_{\rm ren} &=& Z_\phi(D_\mu\phi)^\dagger(D^\mu \phi)
          + Z_\Phi (D_\mu \Phi)^\dagger(D^\mu \Phi)
          - V(\sqrt{Z_\phi}\phi, \sqrt{Z_\Phi}\Phi)
\nonumber \\
  &&~~~+ \sqrt{Z_\Phi}g_t(\overline{\psi}_L t_R \tilde \Phi + {\rm h.c})
          + \sqrt{Z_\phi}g_b(\overline{\psi}_L b_R \phi + {\rm h.c}),
\ee
and matching the lagrangian with Eq. (2.5) 
at the compositeness scale $\Lambda$,
we obtain the matching condition
\be
&& \sqrt{Z_\phi}\rightarrow 1,~~~~~ \sqrt{Z_\Phi} \rightarrow 0,
\nonumber \\
&& Z_\phi \mu_1^2 \rightarrow m_{0}^2,~~~~~
Z_\Phi \mu_2^2 \rightarrow M^2,
\\
&& Z_\phi \lambda_1 \rightarrow \lambda_{10},~~~~~
Z_\Phi^2\lambda_2 \rightarrow 0,
\nonumber \\
&& Z_\phi Z_\Phi \lambda_{i=3,4,5} \rightarrow 0,
\nonumber
\ee
as the scale $\mu \rightarrow \Lambda$.
Now the low energy theory is the Type-II two-Higgs doublet model
with two Higgs fields $\phi$ and $\Phi$ with the compositeness 
conditions for $\Phi$: 
$\Phi$ has a vanishing wavefunction renormalization constant
at the compositeness scale $\Lambda$, while $\phi$ does not.

Before proceeding, we would like to compare our model with Luty's model, 
since both models are two-Higgs doublet models in the low energy regime.
In Luty's model, both Higgs doublets are composite, and thus the matching 
conditions are given by
\be
&& \sqrt{Z_\phi}\rightarrow 0,~~~~~ \sqrt{Z_\Phi} \rightarrow 0,
\nonumber \\
&& Z_\phi \mu_1^2 \rightarrow m_{0}^2,~~~~~
Z_\Phi \mu_2^2 \rightarrow M^2,
\\
&& Z_\phi \lambda_1 \rightarrow 0,~~~~~
Z_\Phi^2\lambda_2 \rightarrow 0,
\nonumber \\
&& Z_\phi Z_\Phi \lambda_{i=3,4,5} \rightarrow 0.
\nonumber
\ee
Namely the conditions for the scalar field $\phi$, the self coupling 
$\lambda_1$ and the wavefunction renormalization constant $Z_\phi$, are
different  from our case. 
These different matching conditions lead to vastly different predictions 
for the scalar boson spectra compared to the Luty's model. 
Also we have additional Yukawa coupling $g_b$ so that we can fit both 
the bottom and the top quark masses without difficulty unlike Luty's model. 

\section{Particle Spectrum}

Our model is defined in terms of three parameters: Higgs self coupling 
$\lambda_{10}$, the compositeness scale $\Lambda$ (where $\lambda_{10}$ and 
the NJL interaction are specified), and the CP-odd Higgs boson mass $m_A$. 
Since $\lambda_{10}$ is also present in the SM, our model has two more 
parameters compared with the SM. 
In order to  study  the low energy phenomenology of our model, 
we relate the model defined at the high scale $\Lambda$ 
to the low-energy spectrum of the theory
by evolving  the renormalization group (RG) equation
from $\Lambda$ to the electroweak scale
with the compositeness conditions, Eq.~(2.9). 
Using the field redefinition
\be
\phi \rightarrow Z_\phi^{-1/2}\phi,~~~~~
\Phi \rightarrow Z_\Phi^{-1/2}\Phi,
\ee
we rewrite the matching condition given in Eq. (2.9) as
\be
&& 
g_b \rightarrow g_{b0}, 
~~~~~ 
g_t \rightarrow \infty,
\nonumber \\
&& \lambda_1/g_b^4 \rightarrow \lambda_{10}/g_{b0}^4, ~~~~~
\lambda_{2,3,4,5} \rightarrow 0,
\ee
for the rescaled couplings.
These conditions are the boundary conditions for the RG equations. 
The condition $ g_t \rightarrow \infty$ and $ \lambda_{2,3,4,5} \rightarrow 0$ 
at the compositeness scale $\Lambda$ are 
the compositeness condition for $\Phi$.
No conditions are assigned on $g_{b0}$ and $\lambda_{10}$ at this stage.
They will be fixed or constrained by the phenomenological conditions 
and the electroweak symmetry breaking  conditions.
We will use the one-loop RG equations given in Ref. \cite{hillleungrao}, 
which are reproduced in the Appendix for convenience.
 
\begin{figure}[t]
\begin{center}
\hbox to\textwidth{\hss\epsfig{file=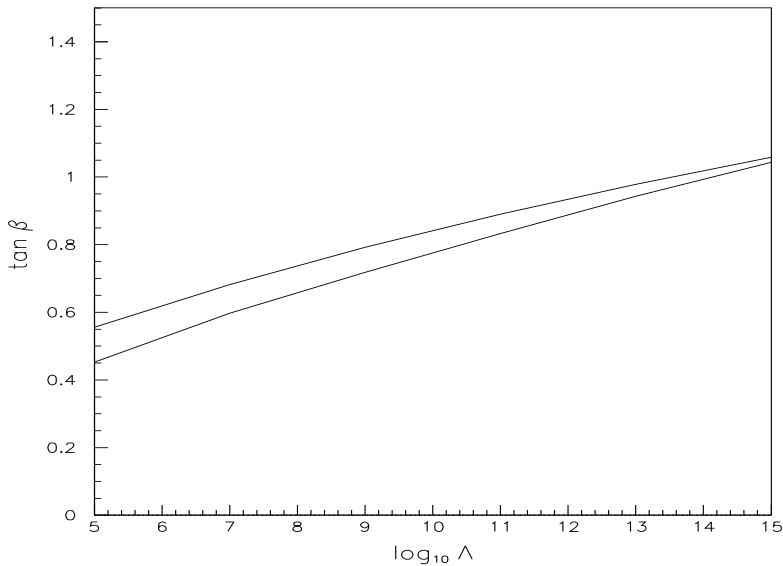,width=12cm,height=9cm}\hss}
\vspace{0.2cm}
\caption{
Allowed values of $\tan \beta$
with respect to the compositeness scale $\Lambda$.
}
\end{center}
\end{figure}

We set the Higgs vacuum expectation values (VEVs) to be
\be
\langle \phi \rangle = \frac{1}{\sqrt{2}}\left(%
\begin{array}{c}
  0 \\
  v_1 \\
\end{array}%
\right),~~~~~
\langle \Phi \rangle = \frac{1}{\sqrt{2}}\left(%
\begin{array}{c}
  0 \\
  v_2 e^{i \delta} \\
\end{array}%
\right),
\ee
with the electroweak symmetry breaking scale
$v^2 = v_1^2 + v_2^2 = (246~{\rm GeV})^2$.
We define $\tan\beta \equiv v_2/v_1$.
The relative phase $\delta$ can be absorbed by an appropriate
field redefinition, and we will drop it hereafter.
The top and bottom quark masses are generated by the VEVs of $\phi$ and 
$\Phi$: 
\begin{eqnarray*}
m_t & = & {1\over \sqrt{2}}~ g_t v_2 = {1\over \sqrt{2}}~g_t v \sin \beta ,
\\
m_b & = & {1\over \sqrt{2}}~ g_b v_1 = {1\over \sqrt{2}}~g_b v \cos \beta .
\end{eqnarray*}
Since the boundary condition for $g_t$ is fixed to be 
the compositeness condition at $\Lambda$, 
$\tan \beta$ is almost fixed in our model.
In the actual numerical analysis, we take the boundary condition of 
$g_t$ to be finite instead of infinity, but large enough for being 
in nonperturbative region, following the approaches of BHL and Luty. 
This approach is supported by the fact that
the low-energy behavior of $g_t$ is not that sensitive to 
the boundary condition at $\Lambda$ when $g_t \to \infty$
because of the infrared fixed point behavior \cite{luty,hillleungrao}.
Once $\tan \beta$ is fixed, the bottom Yukawa coupling $g_{b0}$ at the 
composite scale is also determined by the measured bottom quark mass.
We use the following values for the top and bottom quark masses
\cite{pdg} 
\be
m_t ( M_Z )  = 178.1~{\rm GeV},~~~~~ m_b ( M_Z ) = 2.8~{\rm GeV},
\ee
and gauge couplings as
\be
\alpha (M_Z) = \frac{1}{127.934},~~~
\alpha_s (M_Z) = 0.1172,~~~
\sin^2 \theta_W (M_Z)= 0.2221.
\ee
As a result, only a very narrow window of $\tan \beta$ is consistent
with the measured top and bottom quark masses for a given compositeness 
scale $\Lambda$.  In Fig.~1, we show the allowed range of $\tan\beta$ 
for the compositeness scale $\Lambda$. 


\begin{figure}[t]
\centering
\subfigure[]{
\includegraphics[width=0.65\textwidth]
{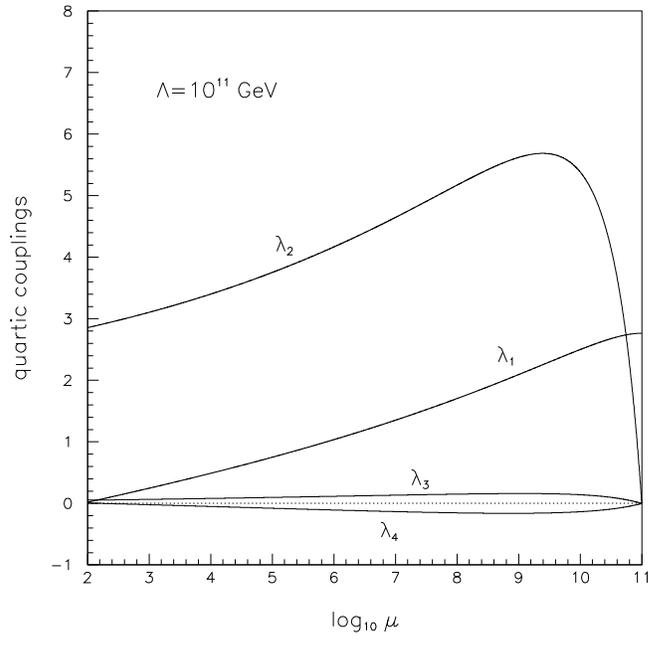}}
\subfigure[]{
\includegraphics[width=0.65\textwidth]
{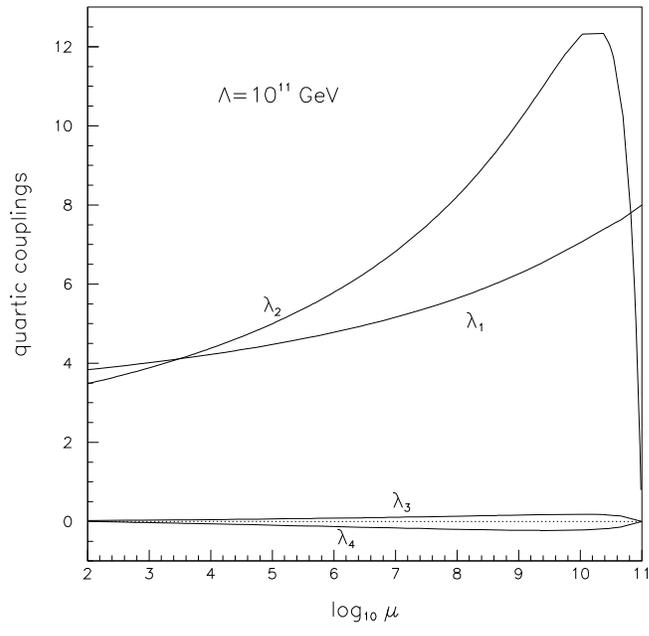}}
\caption{\label{fig:heterom2}
Evolution of the Higgs quartic coupling $\lambda_i$
with respect to $\log_{10} \mu$ for two different $\lambda_{10}$ : 
(a) and (b) . }
\end{figure}

\begin{figure}[t]
\begin{center}
\hbox to\textwidth{\hss\epsfig{file=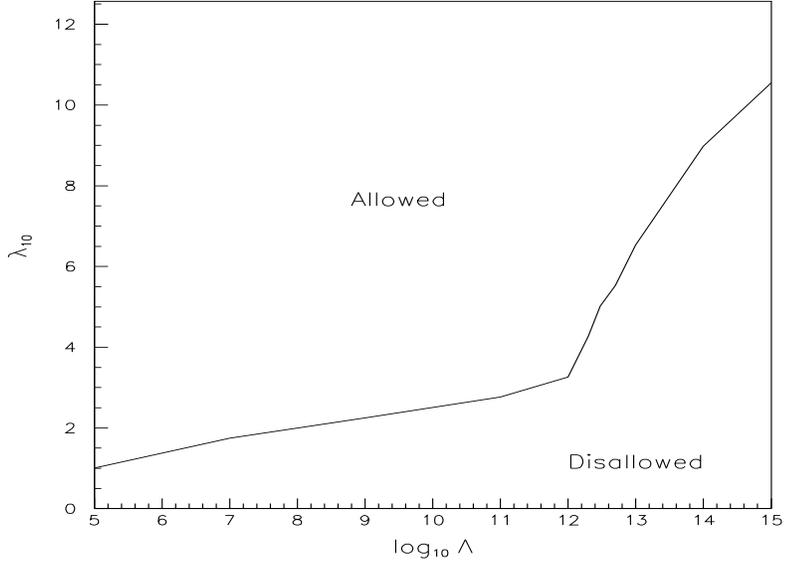,width=12cm,height=9cm}\hss}
\vspace{0.2cm}
\caption{
Allowed values of the quartic coupling $\lambda_{10}$ at $\Lambda$
with respect to the compositeness scale $\Lambda$.
}
\end{center}
\end{figure}

\begin{figure}[ht]
\begin{center}
\hbox to\textwidth{\hss\epsfig{file=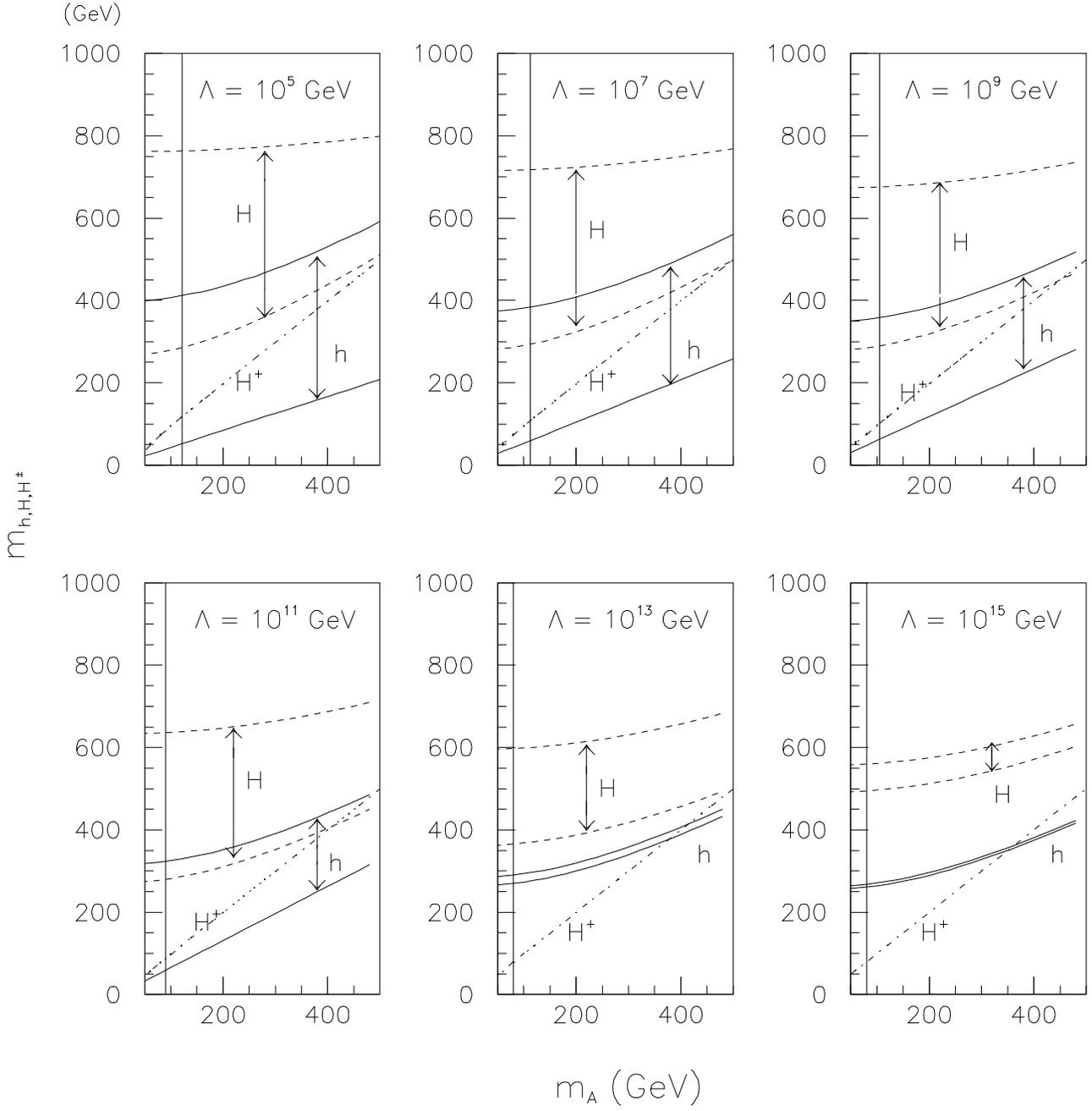,height=18.5cm}\hss}
\vspace{0.2cm}
\caption{
Masses of neutral Higgs bosons $h$ (inside the solid lines), 
$H$ (inside the dashed lines) and the charged Higgs boson $H^\pm$ 
(dahs-dotted line) with respect to $m_A$.
}
\end{center}
\end{figure}

Evolutions of quartic couplings $\lambda_i$ are
constrained by the stability and electroweak symmetry breaking 
conditions of the Higgs potential.
Since the potential should to be bounded from below,
we require that 
\be
\lambda_{1}, ~\lambda_2 & > & 0
\nonumber  \\
\sqrt{\lambda_1 \lambda_2} &>& - \lambda_3 - \lambda_4 + |\lambda_5|
~~~~~{\rm if}~~ \lambda_4 < |\lambda_5|,
\nonumber \\
\sqrt{\lambda_1 \lambda_2} &>& - \lambda_3 
~~~~~~~~~~~~~~~~~~~~{\rm if}~~   \lambda_4 > |\lambda_5| .
\ee
Minimizing the Higgs potential at $v_1$ and $v_2$,
we obtain the following conditions for the electroweak symmetry breaking :
\be
\mu_1^2+\mu_{12}^2 \tan \beta
   + \frac{1}{2}\lambda_1 v_1^2 + \frac{1}{2}\lambda_{345} v_2^2 &=& 0,
\nonumber \\
\mu_2^2+\mu_{12}^2 \cot \beta
   + \frac{1}{2}\lambda_2 v_2^2 + \frac{1}{2}\lambda_{345} v_1^2 &=& 0,
\ee
where $\lambda_{345} = \lambda_3 + \lambda_4 + \lambda_5$.
The boundary condition for $\lambda_1$, $\lambda_{10} \equiv \lambda_1 
( \Lambda) $, is chosen so that the evolved couplings satisfy those 
conditions (3.6) and (3.7) at the electroweak scale.
Typical evolutions of quartic couplings for two different $\lambda_{10}$ 
are shown in the Fig. 2 (a) and (b).
Since $\lambda_1$ decreases almost monotonically,
it becomes negative at the electroweak scale in most region
of the parameter space when $\Lambda > 10^{12}$ GeV if $\lambda_{10}$ is 
small or moderate [ Fig.~2 (a) ]. 
For larger $\lambda_{10}$, the RG evolved $\lambda_1$ could
be positive at electroweak scale [ Fig.~2 (b) ]. In such a case, the 
scalar self coupling $\lambda_{1,2}$ can be large so that the resulting
triple or quartic Higgs self coulings could be enhanced significantly 
compared to the SM case. 
After all, the allowed parameter space for $\lambda_{10}$ is significantly  
reduced for large compositeness scale $\Lambda$.  
The allowed region of $\lambda_{10}$ is shown in Fig.~3 for 
different values of the compositeness scale $\Lambda$.

The coupling $\lambda_5$ is very interesting, since it can contribute 
to the CP violation in the Higgs sector through mixing, 
if ${\rm Im} ( \lambda_5 ) \neq 0$. 
It turns out that the one-loop beta function for $\lambda_5$ is proportional 
to $\lambda_5$ itself [ see Eq. (5.4) ]. Since we have the vanishing initial 
condition for $\lambda_5 = 0$, we have $\lambda_5 = 0$ at all the scale 
down to $M_Z$. 
The $\mu_{12}^2$ term can also contribute to the CP violation in the Higgs 
sector if it has a CP violating  phase. However the minimization condition
for the scalar potential leads to \cite{mu12} 
\[
{\rm Im} ( \mu_{12}^2 ) = - {1\over 2} ~{\rm Im} ( \lambda_5 ) v_1 v_2 ,
\]
which can be obtained from the imaginary part of (3.6).
Since $\lambda_5 = 0$ in our model, we have ${\rm Im} ( \mu_{12}^2 ) = 0$ 
and no Higgs sector CP violation from the phase of $\mu_{12}^2$ term 
in our model.  This gurantees that there will be no mixing between the 
CP-even and the CP-odd neutral Higgs bosons. 
Finally, we will demand that the perturbativity condition
$\lambda < 4 \pi$ be satisfied in this analysis.

\begin{figure}[ht]
\begin{center}
\mbox{\epsfig{file=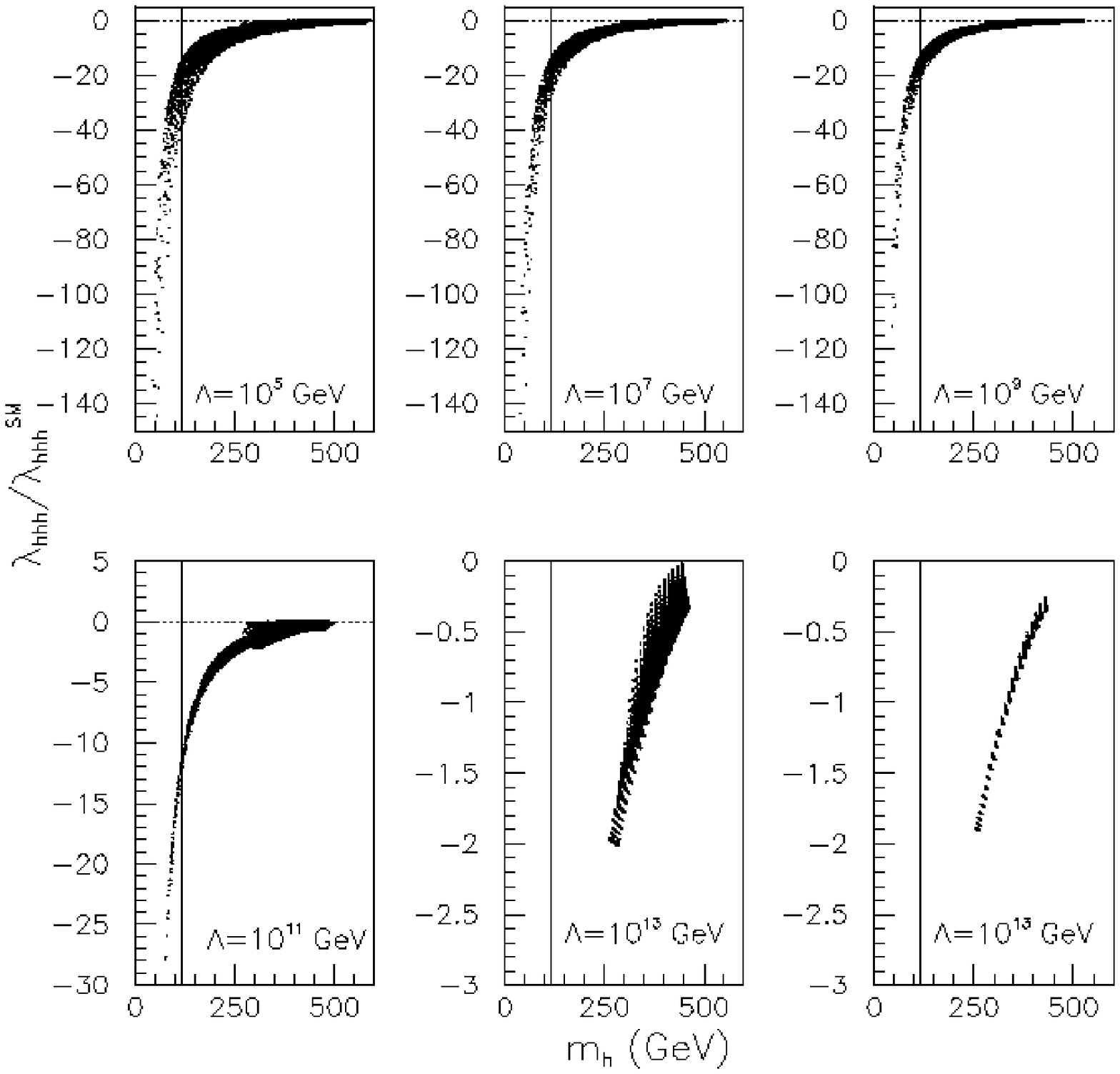,width=15.0cm}}
\vspace{0.2cm}
\caption{
Ratio of our model prediction for the triple self-coupling 
of the Higgs boson to that of the SM with respect to the lightest
neutral Higgs boson mass. The vertical line denotes the lower bound
on the Higgs boson mass
}
\end{center}
\end{figure}

After EWSB, we have three neutral scalar fields
and a couple of charged scalar.
The mass matrix for real part of the neutral scalar fields 
$({\rm Re} \phi, {\rm Re} \Phi)$ is given by
\be
M^2 = \left(
\begin{array}{cc}
  \mu_1^2 + \frac{3}{2}\lambda_1 v_1^2 + \frac{1}{2} v_1 v_2 \lambda_{345}
  & \mu_{12}^2 + v_1 v_2 \lambda_{345} \\
  & \\
  \mu_{12}^2 + v_1 v_2 \lambda_{345}
  & \mu_2^2 + \frac{3}{2}\lambda_2 v_2^2 + \frac{1}{2} v_1 v_2 \lambda_{345}
   \\
\end{array}%
\right).
\ee
Diagonalizing the mass matrix,
the physical CP-even Higgs bosons are defined by
\be
H &=& \sqrt{2}~( {\rm Re} \Phi \sin \alpha + {\rm Re} \phi \cos \alpha ),
\nonumber \\
h &=& \sqrt{2}~( {\rm Re} \Phi \cos \alpha - {\rm Re} \phi \sin \alpha ),
\ee
with the mixing angle $\alpha$ defined by
\be
\tan 2 \alpha \equiv \frac{2 M_{12}^2}{M_{11}^2 - M_{22}^2},
\ee
and their masses are given by
\be
m_{H,h} = \frac{1}{2} \left[ M_{11}^2 + M_{22}^2 \pm
            \sqrt{ (M_{11}^2 - M_{22}^2)^2 + 4 M_{12}^2 } \right].
\ee

\begin{figure}[ht]
\begin{center}
\mbox{\epsfig{file=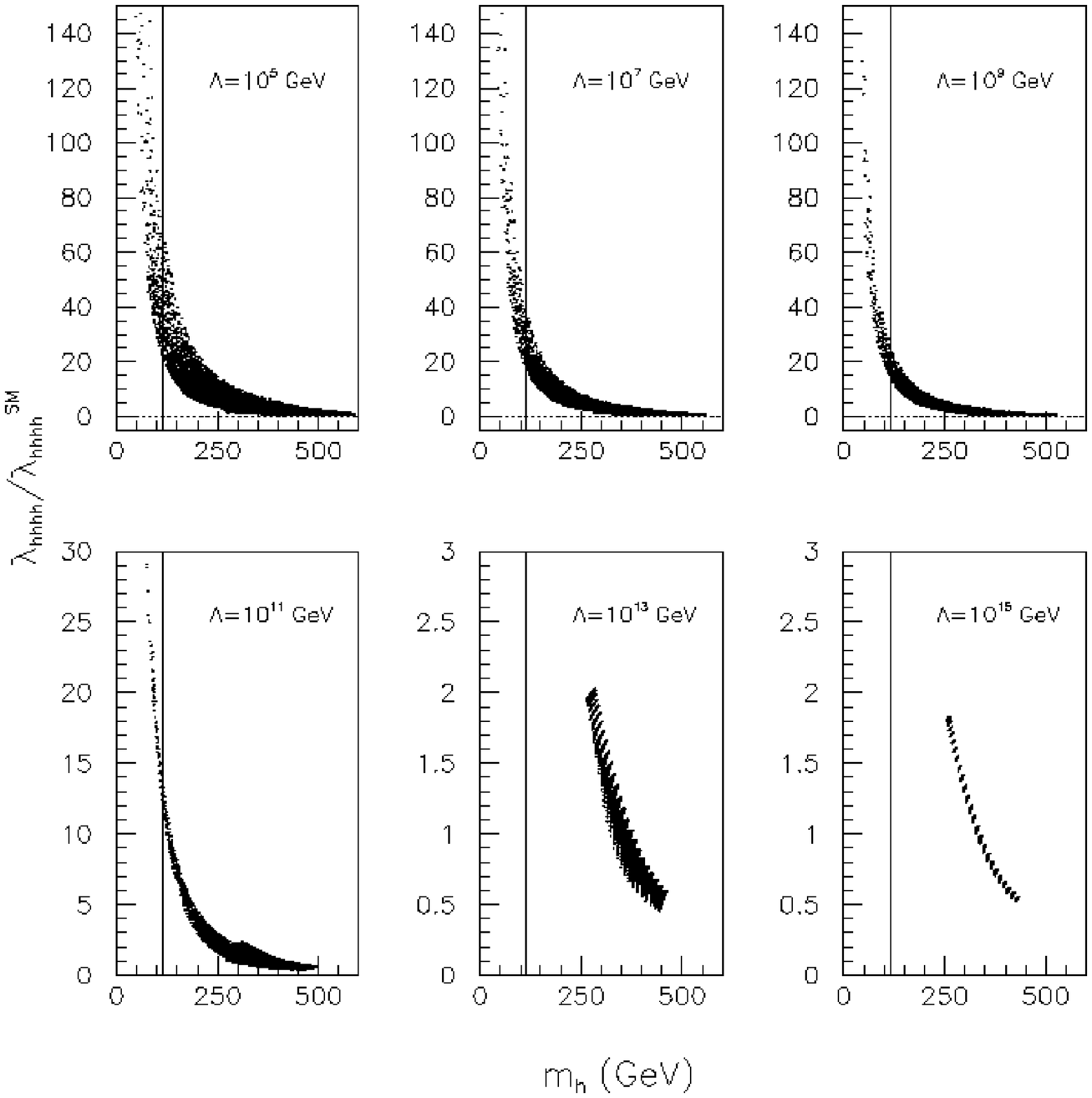,width=15.0cm}}
\vspace{0.2cm}
\caption{
Ratio of our model prediction for the quartic self-coupling 
of the Higgs boson to that of the SM with respect to the lightest
neutral Higgs boson mass. The vertical line denotes the lower bound
on the Higgs boson mass
}
\end{center}
\end{figure}

The CP-odd Higgs boson $A$ is defined by the orthogonal state
of the pseudoscalar Goldstone mode,
$A = \sqrt{2} (-{\rm Im} \Phi \sin \beta + {\rm Im} \phi \cos \beta)$
and its mass is given by
\be
m_A^2 = -\frac{2 \mu_{12}^2}{\sin 2 \beta} - \lambda_5 v^2.
\ee
Since $\lambda_5 = 0$ in our model, the mass of the CP-odd Higgs boson 
$A$ is generated solely by the soft breaking term $\mu_{12}^2$ whch is 
another free parameter in our model.
Therefore, we take the mass of the CP-odd Higgs boson $m_A$
to be an input parameter in our analysis, traded with $\mu_{12}^2$, 
as mentioned earlier.

Finally, the charged Higgs boson masses are given by
\be
m_{H^\pm}^2 = m_A^2 - \frac{1}{2} (\lambda_4 -\lambda_5) v^2 = 
m_A^2 - \frac{1}{2}~\lambda_4 v^2.
\ee
The masses of neutral and charged Higgs bosons with varying $m_A$ 
are given in Fig.~4.
The vertical line denotes the mass bound of $A$
derived from the bound on $m_{H^\pm}$ as given in PDG. 
Generically experimental bound on $m_{H^\pm}$ is more stringent than 
that on $m_A$.
We note that $\lambda_4$ becomes positive at the electroweak scale
in most cases unlike Luty's model, since the contributions of 
$\lambda_1$ and $\lambda_2$ to the evolution is substantial.
Consequently, the charged Higgs boson $H^{\pm}$ is generically lighter 
than the CP-odd Higgs boson in our model.
As shown in Fig.~4, moreover, $H^\pm$ may be even lighter
than the lightest neutral Higgs boson when the compositeness scale 
$\Lambda$ is high. This is a generic feature of our model.

\section{Low Energy Phenomenology}

\subsection{Higgs boson self couplings}

The Higgs boson self-couplings may play a role of the sensitive probe
to the new physics and have great phenomenological significance 
\cite{selfcoupling,selfcoupling1}.
The ratios of the triple and the quartic self-couplings of the lighter 
CP-even Higgs boson $h$ in general two Higgs doublet model
to those of the SM are given by \cite{selfcoupling1} 
%
The ratio of self-couplings to the SM values are given by
\be
\frac{\lambda_{hhh}}{\lambda^{\rm SM}_{hhh}}
&=& \frac{1}{4 \sqrt{2}} \left(
-\frac{\lambda_{1}}{\lambda_{\rm SM}} \cos \beta \sin^3 \alpha
+\frac{\lambda_{2}}{\lambda_{\rm SM}} \sin \beta \cos^3 \alpha
-\frac{\lambda_{345}}{\lambda_{\rm SM}}
\cos \alpha \sin \alpha \cos (\alpha+\beta)
\right),
\nonumber \\
\frac{\lambda_{hhhh}}{\lambda^{\rm SM}_{hhhh}}
&=& \frac{1}{8} \left(
\frac{\lambda_{1}}{\lambda_{\rm SM}} \sin^4 \alpha
+\frac{\lambda_{2}}{\lambda_{\rm SM}} \cos^4 \alpha
+2\frac{\lambda_{345}}{\lambda_{\rm SM}}
\cos^2 \alpha \sin^2 \alpha \right),
\ee
where $\alpha$ is the mixing angle defined in (3.9) and (3.10). 
When $\tan \alpha$ tends to be very large,
$\cos \alpha \ll \sin \alpha$ and the first terms of the
above equations dominate in most parameter space.
Thus $\lambda_1/\lambda_{\rm SM}$ plays the crucial role
for the ratio of self-couplings. And this could be large if the initial
$\lambda_{10}$ is large, as shown in Fig.~2 (b).   
In Fig.~5 and Fig.~6, we  show $\lambda_{hhh}/\lambda_{hhh}^{\rm SM}$ and  
$\lambda_{hhhh} / \lambda_{hhhh}^{\rm SM}$ as functions of $m_h$. 
We find that the deviations of the self-couplings from the SM values 
can be substantial, even close to 20 times, when the CP-even Higgs boson
$h$ becomes light.
For the triple self-coupling, such an enhancement arises
when $m_A > m_h$ in the small $m_h$ region
through the squared mass ratio $m_A^2/m_h^2 \sim 20$
in the second term of Eq. (4.1).
Thus our triple self-coupling is negative in most parameter region,
which can lead to many interesting phenomenologies
due to the destructive interference with the SM contribution.
For the quartic self-coupling, the ratio of $m_H^2/m_h^2$
is responsible for the enhancement since $m_H > m_A$
in most region of parameter space.
These large deviations of Higgs self couplings from the SM predictions 
are the most interesting features of our model with one fundamental and 
one composite Higgs fields.
The Higgs pair and the triple Higgs production cross sections at the ILC
will be interesting signatures of our model, and the detailed 
phenomenology will be discussed elsewhere \cite{work1}. 

\subsection{Signatures at the LHC}

The main production mechanism for the lighter neutral CP-even Higgs boson 
$h$ at the LHC is the gluon-gluon fusion process, $gg \to h$.
The top quark loop determines this process both in the SM and in our 
model since $\tan \beta \lesssim O(1)$. The cross section predicted in our 
model is enhanced from that in the SM by a factor \cite{djouadi},
\be
\sigma (pp \rightarrow gg \rightarrow h^0 ) 
= \left( \frac{\cos \alpha}{\sin \beta} \right)^2 
\sigma_{SM}(pp \rightarrow gg \rightarrow h^0 ),
\ee
where the mixing angle $\alpha$ is given in Eq. (3.10). Our model
 prediction is depicted in Fig.~7.
Moreover, there are more Higgs bosons in our model compared 
with the SM. The CP-odd Higgs boson $A$ can be lighter than the CP-even
Higgs (see Fig.~4) and then it is possible that the first observed 
Higgs would  not be the CP-even $h$ and but the CP-odd $A$.
On the other hand, the charged Higgs boson $H^{\pm}$ can be lighter than $h$
(see Fig.~4).  When the compositeness scale $\Lambda$ is high enough
and $m_h < 350$ GeV, $m_h$ is always larger than $m_{H^\pm}$.
In that case, the charged Higgs could be observed before the neutral
Higgs boson through the $g b$ fusion. 

\subsection{Signatures at the ILC}

The future $e^+ e^-$ international linear collider (ILC) will examine
the detailed structure of the Higgs sector (or EWSB sector).
The most important channel for the neutral Higgs boson production 
at the ILC are the Higgs-strahlung process ( $e^- e^+ \to Z H$ )
and the $WW$ fusion process 
( $e^- e^+ \to W^\ast W^\ast \to \bar{\nu}_e \nu_e H$ ) \cite{djouadi}
where $H$ is one of the CP-even Higgs bosons $ h^0, H^0$.
The cross sections for these processes are expressed by
\be
\sigma (e^+ e^- \to Z + h^0 / H^0) &=&
\sin^2 ~/~ \cos^2 (\beta - \alpha)
\sigma_{\rm SM} (e^+ e^- \to Z + h^0 / H^0),
\nonumber \\
\sigma (e^- e^+ \to \bar{\nu}_e \nu_e + h^0/H^0)
&=& \sin^2 ~/~ \cos^2 (\beta - \alpha)
\sigma_{\rm SM} (e^- e^+ \to \bar{\nu}_e \nu_e + h^0/H^0).
\ee
The SM cross section $\sigma_{\rm SM}$ is given in Ref. \cite{djouadi}.
In Fig.~7, we show  the Higgs production cross sections as functions 
of $m_h$. Depending on the neutral Higgs mixing angle $\alpha$,
the predicted cross section can take a wide range compared to the 
SM predition, especially for low compositeness scale $\Lambda$. 
For higher $\Lambda$, the allowed $m_h$ has a narrow region, 
since the input parameter $\lambda_{10}$ and the mixing angle $\alpha$ are 
strongly constrained if the compositeness scale $\Lambda$ is high. 
Therefore the Higgs production cross section is almost definitely 
determined as a function of $m_h$,  when $\Lambda > 10^{12}$ GeV.

\begin{figure}[ht]
\begin{center}
\hbox to\textwidth{\hss\epsfig{file=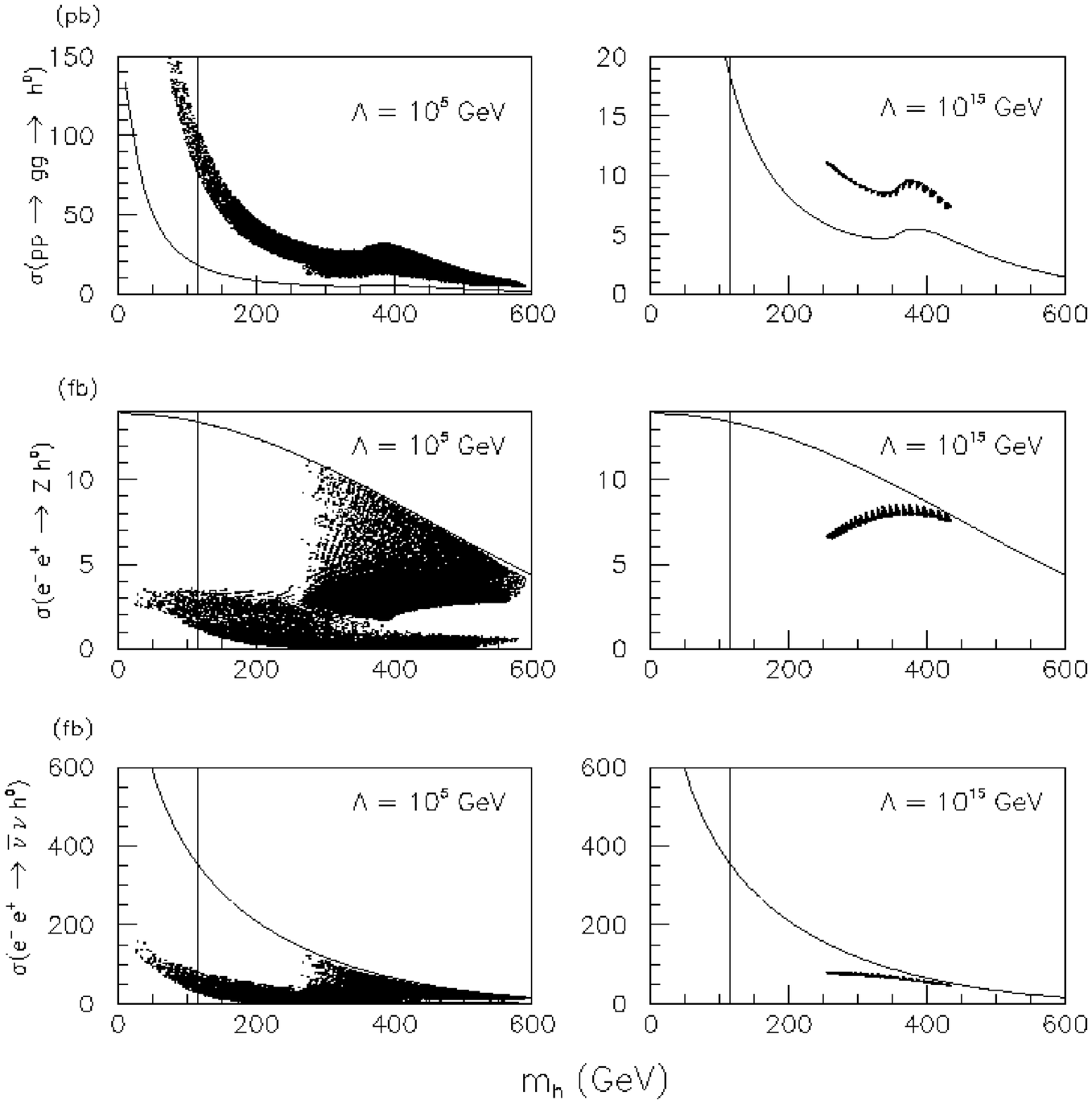,width=17cm}\hss}
\vspace{0.2cm}
\caption{
Production  cross section of the neutral Higgs boson at 
the LHC and ILC. $\sqrt{s} = 14$ TeV for the LHC and  $\sqrt{s} = 1$ TeV for
the ILC are assumed.
The solid curves denotes the SM predictions.
}
\end{center} 
\end{figure}

\section{Conclusions}

In conclusion, we considered an interesting possibility that the Higgs 
boson produced at the future colliders is neither a fundamental scalar
nor a composite scalar, but a mixed state of them.
It could be a  generic feature, if there exists a strong dynamics
at a high scale which give rise to the dynamical electroweak symmetry 
breaking, in addition to the usual Higgs mechanism due to the 
nonvanishing  VEV  of a fundamental Higgs. 
It is interesting that this scenario could be easily realized,  
if we embed the SM lagrangian in a higher dimension with bulk gauge 
interactions.
The bulk gauge interaction can give rise to a new strong dynamics
in the 4 dimensional theory and trigger the dynamical symmetry
breaking of the electroweak symmetry.
We have constructed the simplest model with the NJL type four-fermion
interaction of top quarks as the strong dynamics inspired by the BHL
and study the phenomenology of the two Higgs doublets model 
with the compositeness condition as the low energy effective theory.
The resulting theory can accommodate the observed top mass, and give 
specific predictions for neutral and charged Higgs masses at a given value
of $\Lambda$. For example, the charged Higgs boson is always lighter than
the CP-odd Higgs neutral boson, although the mass difference is very small.
For $\Lambda \sim 10^{15}$ GeV, the allowed parameter region
is rather restricted, and we predict $m_h > 250$ GeV. 
Also the charged Higgs boson  becomes lighter than $h$, and 
could be the first signal of our model at the future colliders. 

Our study can be extended into various directions. One can consider more 
general situation where the fundamental Higgs couples both to the bottom 
and the top quarks. In that case, the resulting low energy effective 
theory will be a Type-III two-Higgs doublet model. Or one can consider
both $t\bar{t}$ and $b\bar{b}$ condense and contribute to the EWSB.
Then the resuting theory will be a three-Higgs doublet model with specific
matching conditions.  Also one has to include all the three generations 
and construct realistic models with correct CKM mixing and CP violation. 
Some of these issues will be pursued in separate publications in the future.

\bigskip


\acknowledgments
PK is grateful to Bogdan Dobrescu and Chris Hill for useful discussions.
This work is supported in part by KRF Sundo grant R02-2003-000-10085-0,
and KOSEF through CHEP at Kyungpook National University (PK),
and by Korea Research Foundation Grant KRF-2003-050-C00003 (KYL).

\section*{Appendix : RG equations}

In this work, we use the 1-loop RG equations for the Type-II Higgs 
doublet model as given in Ref. \cite{hillleungrao}.
The RG equations for gauge couplings are given by
\be
Dg_i = - c_i g_i^3,
\ee
where $D = 16 \pi^2 \partial / \partial \ln \mu$ and
\be
c_1 = -\frac{1}{6}N_H-\frac{20}{9}N_g,~~~
c_2 = \frac{22}{3}-\frac{4}{3}N_g-\frac{1}{6} N_H,~~~
c_3 = 11 - \frac{4}{3}N_g,
\ee
with the number of generations $N_g$ and
the number of Higgs doublets $N_H$.

The RG equations for Yukawa couplings are given by
\be
Dg_b = - (8g_3^2 + \frac{9}{4}g_2^2+\frac{5}{12}g_1^2)g_b + \frac{9}{2}g_b^3
         +\frac{1}{2}g_t^2g_b,
\nonumber \\
Dg_t = - (8g_3^2 + \frac{9}{4}g_2^2+\frac{17}{12}g_1^2)g_t + \frac{9}{2}g_t^3
         +\frac{1}{2}g_b^2g_t,
\ee
and for Higgs quartic self-couplings given by
\be
D\lambda_1 = && 12\lambda_1^2+4\lambda_3^2+4\lambda_3\lambda_4+2\lambda_4^2
             +2\lambda_5^2-3\lambda_1(3g_2^3+g_1^2)+\frac{3}{2}g_2^4
\nonumber \\
            &&+\frac{3}{4}(g_2^2+g_1^2)^2 +12\lambda_2g_b^2-12g_b^4
\nonumber \\
D\lambda_2 = && 12\lambda_1^2+4\lambda_3^2+4\lambda_3\lambda_4+2\lambda_4^2
             +2\lambda_5^2-3\lambda_2(3g_2^3+g_1^2)+\frac{3}{2}g_2^4
\nonumber \\
            &&+\frac{3}{4}(g_2^2+g_1^2)^2 +12\lambda_1g_t^2-12g_t^4
\nonumber \\
D\lambda_3 = && (\lambda_1+\lambda_2)(6\lambda_3+\lambda_4)+4\lambda_3^2
             +2\lambda_4^2+2\lambda_5^2-3\lambda_3(3g_2^2+g_1^2)
\nonumber \\
             &&+\frac{9}{4}g_2^4+\frac{3}{4}g_1^4-\frac{3}{2}g_2^2g_1^2
             +6\lambda_3(g_b^2+g_t^2)-12g_b^2g_t^2
\nonumber \\
D\lambda_4 = && (\lambda_1+\lambda_2)\lambda_4
             +4(2\lambda_3+\lambda_4)\lambda_4+8\lambda_5^2
             -3\lambda_4(3g_2^2+g_1^2)
\nonumber \\
             &&+3g_1^2g_2^2+6\lambda_4(g_b^2+g_t^2)
             +12g_b^2g_t^2
\nonumber \\
D\lambda_5 = &&\lambda_5\left[2(\lambda_1+\lambda_2)+8\lambda_3+12\lambda_4
             -3(3g_2^2+g_1^2)+6(g_b^2+g_t^2)\right].
\ee

%

\def\PRD #1 #2 #3 {Phys. Rev. D {\bf#1},\ #2 (#3)}
\def\PRL #1 #2 #3 {Phys. Rev. Lett. {\bf#1},\ #2 (#3)}
\def\PLB #1 #2 #3 {Phys. Lett. B {\bf#1},\ #2 (#3)}
\def\NPB #1 #2 #3 {Nucl. Phys. {\bf B#1},\ #2 (#3)}
\def\ZPC #1 #2 #3 {Z. Phys. C {\bf#1},\ #2 (#3)}
\def\EPJ #1 #2 #3 {Euro. Phys. J. C {\bf#1},\ #2 (#3)}
\def\JHEP #1 #2 #3 {JHEP {\bf#1},\ #2 (#3)}
\def\IJMP #1 #2 #3 {Int. J. Mod. Phys. A {\bf#1},\ #2 (#3)}
\def\MPL #1 #2 #3 {Mod. Phys. Lett. A {\bf#1},\ #2 (#3)}
\def\PTP #1 #2 #3 {Prog. Theor. Phys. {\bf#1},\ #2 (#3)}
\def\PR #1 #2 #3 {Phys. Rep. {\bf#1},\ #2 (#3)}
\def\RMP #1 #2 #3 {Rev. Mod. Phys. {\bf#1},\ #2 (#3)}
\def\PRold #1 #2 #3 {Phys. Rev. {\bf#1},\ #2 (#3)}
\def\IBID #1 #2 #3 {{\it ibid.} {\bf#1},\ #2 (#3)}

\end{document}